# Dynamic Performance Management: An Approach for Managing the Common Goods

**Alberto Sardi * and Enrico Sorano**

Department of Management, University of Torino, 10124 Torino, Italy; alberto.sardi@unito.it (A.S.); enrico.sorano@unito.it (E.S.)
*   Correspondence: alberto.sardi@unito.it

**Abstract:** Public organizations need innovative approaches for managing common goods and to explain the dynamics linking the (re)generation of common goods and organizational performance. Although system dynamics is recognised as a useful approach for managing common goods, public organizations rarely adopt the system dynamics for this goal. The paper aims to review the literature on the system dynamics and its recent application, known as dynamic performance management, to highlight the state of the art and future opportunities on the management of common goods. The authors analyzed 144 documents using a systematic literature review. The results obtained outline a fair number of documents, countries and journals involving the study of system dynamics, but do not cover sufficient research on the linking between the (re)generation of common goods and organizational performance. This paper outlines academic and practical contributions. Firstly, it contributes to the theory of common goods. It provides insight for linking the management of common goods and organizational performance through the use of dynamic performance management approach. Furthermore, it shows scholars the main research opportunities. Secondly, it indicates to practitioners the documents providing useful ideas on the adoption of system dynamics for managing common goods.

## 1. Introduction

Organizations need innovative approaches for managing common goods and linking the (re)generation of common goods and organizational performance management [1]. In fact, the common good can lose its value, for instance, through the users' behaviour (e.g., beneficiaries' neglect, mistakes, or disorganization) [1]. In this sense, the users have to contribute not only to the generation of common goods, but also their regeneration; otherwise, they lose value. We here define this concept with the term "(re)generation" [1].

The development of innovative approaches is crucial, in particular for public organizations, because two of the main general purposes of a public organization are usually a) the management of public goods, and b) the sustainability of the 'public system', as outlined by new public management reform [2]. Although numerous public organizations implemented managerial models traditionally used by non-public organizations, for example, Balance Scorecard, these models rarely represented the managerial complexity of a public organization, thereby failing in their purpose [3–5]. In this sense, system dynamics may be a useful approach for managing this linking in the public context [6,7]. Since 1970, system dynamics has been recognised as a useful approach to study the complexity of a public context in terms of economic, environmental and social changes [8]. In particular, its recent application, known as dynamic performance management, can better support the management of common goods and organizational performance [1,9]. It is an adaptive, feedback-based and learning-oriented performance management system. It allows identifying, mapping and operationalizing feedback loops between variables, such as vicious and virtuous cycles [7,10,11]. However, although dynamic performance management is recognised as an innovative approach to linking the (re)generation of common goods and organizational performance, it is still not adopted widely enough for this purpose [7,10–13].

The aim here is to (a) review the literature on system dynamics and its recent application, dynamic performance management and (b) highlight the main future opportunities of its use in the public sector. For (a), we show the number of documents, countries and journals involved in the research of dynamic performance management. Then, we describe the topics, the models and the common goods analyzed for each paper detected. After these analyses, for (b) we suggest the main future opportunities for adopting this innovative approach of the management of common goods.

Using a rigorous systematic literature review as suggested by Tranfield et al., we analyzed 144 documents related to dynamic performance management in the public context. The documents involved cover the period from 1976 to 2018.

The main findings obtained describe a fair number of documents, countries and journals involving the research of dynamic performance management, but do not cover sufficient studies in system dynamics adopted for managing common goods according to the last managerial definition of common goods [1].

The next section outlines, in brief, the background literature on common goods and organizational performance management. Section three describes the methodology adopted to produce the systematic review literature. Section four shows the main findings of this analysis. Section five discusses the state of the art and future opportunities in this area. Finally, the last section summarizes the main findings and highlights the research limitations and future opportunities.

## 2. Literature Background

In order to support a better understanding of this paper, this section highlights the main concepts of this research, that is, common goods and performance management.

### *2.1. Common Goods*

What are common goods? What are the differences between public goods and common goods? What are the main gaps in managing common goods? The answers to these questions can help to understand better common goods.

According to the management perspective, we define a common good as "a resource that is available for collective use by a community, and whose value and availability can only be maintained and/or developed through the collaboration of the beneficiaries" [1], even if it differs by the traditional interpretations, that is, "a common good is a resource that is available for collective use" [14]. In the last decade, the scientific debate on this topic has increased the interpretations of what is a common good [15,16]. The above definition, however, is the last theorized in the literature and the widest interpretation between all the definitions. In the following, we explain better what the differences are between public goods and common goods, why the definition of common goods has changed in the last decade and what the main gaps are for managing common goods.

The general meaning of common goods differs from public goods because public goods are defined as a resource that is available for collective use but not "rivalrous", whilst the common good is defined as a resource that is available for collective use but "rivalrous" [17]. The public good is "non-rivalrous" because the cost of providing it to an additional consumer is zero. The common good is "rivalrous" because its consumption by one consumer prevents or reduces its consumption by other consumers [17]. From the management perspective, the last definition includes the concepts of value and availability [18,19]. For instance, the value of Wikipedia, that is, a common good, is the quality of content; whereas the availability regards its contents online [1]. The common good can also lose its value through the users' behaviour (e.g., beneficiaries' neglect, mistakes, or disorganization) [1,20]. For instance, the users of Wikipedia have to contribute to its contents, otherwise it loses value.

In conclusion, the ability to (re)generate available value depends on its ecosystems (with economic, environmental, social, and technological components) [20,21]. For instance, Wikipedia is an integrated ecosystem able to (re)generate an important common good—its contents [22]. By its definition, Wikipedia is a multilingual online encyclopedia, created and maintained as an open collaboration project using a wiki-based editing system [3].

There are many organizational success factors for managing common goods for instance, regulation, learning and training, decision-making processes [23]. Although there are many organizational success factors for managing common goods, the link between the (re)generation of common goods and organizational performance seems to be one of the most important [1]. However, this link is often relegated to the study of processes which corporate social sustainability [24–28] and sustainability reporting [29–32], rather than performance management. As outlined by Ricciardi et al., "The common goods should be the final goal of organizational performance" and, in this sense, the management of common good should be managed by organizations in order to obtain a competitive advantage. However, this process still lacks dynamic approaches that link organizational learning, organizational performance, and common goods [1].

The literature lacks innovative approaches for linking the management of common goods and organizational performance [1,13,33,34], even though the final goals of the public organization should be the protection and regeneration of common goods [1].

By this paper, we suggest the use of system dynamics to address this gap to contribute to the theory on common goods providing a useful approach for managing common goods and linking common goods and organizational performance management in the public sector.

*2.2. Performance Management*

Performance measurement and management can be the key processes for linking common goods and organizational performance [1]. The performance measurement process is defined as "what to measure" and includes technical activities such as collecting, analyzing and reporting [35], whereas the performance management process is defined as "how to manage performance" and includes social activities such as communicating, rewarding, learning and improving [36,37]. The balance between performance measurement and management establishes the roots of performance measurement systems, that is, the reporting process that gives feedback to employees on the outcome of actions [38,39].

The literature highlights an increasing amount of research and citations on performance measurement and management of public organizations in the last 40 years [39]. On the one hand, the main research on performance measurement activities deals with setting performance measures [40], the design of a benchmarking system and the definition of key performance indicators [41]. However, the most important publications deal with the design of the Balanced Scorecard [42,43].

On the other hand, the main research on performance management activities deals with internal and external communication of performance information [44,45], adoption of the Balanced Scorecard [46] and the use of performance measurement system and reports [47,48].

Although performance management is one of the most pressing challenges for scholars, practitioners and governments, the literature rarely highlights the implementation of performance management achieving expected results.

Recent research projects point to the need for a holistic strategic management approach to link strategic planning to performance management [49]. This type of new approach, also referred to as whole-of-government [50], rarely includes economic, environmental and social factors. There is a need for rethinking how scholars research the field of performance measurement and management by holistic approaches, recognizing the integrated and concurrent nature of challenges [39]. Although recent publications highlight an interest in performance management by public organizations, especially in the educational and healthcare sectors, too often the literature puts the accent on the performance measurement aspect or the unsuccessful performance management model.

We here argue that a dynamic performance management approach can contribute to filling these two gaps:

a) The lack of holistic approaches to performance management still presents a great challenge. Specifically, the lack of models for including the complexity of a public context [3,39,51].
b) The lack of approaches for linking the management of common goods and organizational performance [1,13,33], even though the final goal of a public organization should be the (re)generation of common goods.

We suggest that a dynamic performance management approach can contribute to measure and manage the resources available for collective benefit, and the system's capacity of (re)generation of its own resources [52]. It can also contribute to the inclusion of more actors in public policy analysis [53,54] and in forecasting behaviour and policy at each stage of development. Policymakers may have available a more integrated, transparent and holistic view of the link between various factors [55].

## 3. Methodology

The research project was based on a systematic literature review, differing from the traditional narrative literature review in that it involves a more explicit selection process for documents [56,57], allowing critical appraisal of the relevant primary research in order to highlight the literature review on dynamic performance management in the public context.

To make the literature review replicable, scientific and transparent, we made use of the five stages of the systematic literature review suggested by Tranfield et al.

### 3.1. Stage 1: Planning the Review and Identifying Keywords

In stage one, we mapped the field of study, reviewing key papers and interviewing academics, practitioners and consultants active in the field to define the keywords for the review process [56,57].

To identify a field/sub-field of study we made informal consultations with academics working in the field, practitioners working in the field and information scientists [56,57]. The keywords identified during the consultations included "public", "system dynamics" and "dynamic performance management". The keyword "public" comprises keywords such as "public policies", "public administration", "public organization", "public sector", "public context" and others. In turn, "system dynamics" represents the approach investigated, whereas "dynamic performance management" describes a specific system dynamics approach applied to performance management.

### 3.2. Stage 2: Identifying the Research Criteria

In stage two, we examined the peer-reviewed literature available on Scopus and Web of Science because the Scopus database has the best coverage in the field and Web of Science has the best complementary database [57]. The search included papers published from 1976 (date of the first published paper) to 2019 (10 January). Keywords sought were restricted to abstract, titled and keywords. For research, preference was given to English language documents (Table 1) published in scientific journals covering the fields under study, that is, "Business, Management and Accounting", "Economics, Econometrics and Finance" and "Decision Science". The 176 documents selected were processed using Mendeley software.

**Table 1.** Research Criteria.

| Dataset | Elsevier's Scopus and Web of Science |
|---|---|
| Time | From 1976 (date of the first document published on Scopus) to 2019 |
| Source | Titled, abstract, keyword |
| Document type | Article |
| Source type | Journals |
| Subject area | "Business, Management and Accounting", "Economics, Econometrics and Finance" and "Decision Sciences" |
| Language | English |
| Keywords | "Public" and "Dynamic performance management" or "System dynamics" |

*3.3. Stage 3: Extracting the Relevant Documents*

In stage three, two researchers independently assessed the relevance of 176 documents by reading the abstract. After discarding non-relevant contributions, papers were read fully and assessed. A total of 144 documents were considered relevant for the aim of the research (Table 2).

**Table 2.** Selection process of relevant documents.

| Step | Result |
|---|---|
| Database Searches — Select papers from Elsevier's Scopus | 176 Documents |
| Read Abstract — Eliminate non-relevant papers by reading abstracts | 162 Documents |
| Read Full Papers — Eliminate non-relevant full papers | 144 Documents |

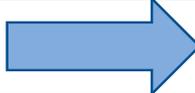
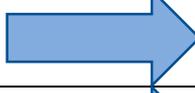
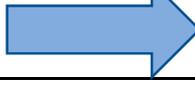

*3.4. Stage 4: Reporting of Main Information*

In stage four, we reported the information in a data extraction sheet format, including all 144 documents and highlighting the following information: (a) publication details—Title, Year, Journal, Author/s, Citations, Keywords; (b) theoretical/empirical (if empirical, countries of analysis); and (c) the topic and model developed for each paper. In order to identify the topic and model developed for each paper, we read the full papers. After that, we categorized the documents. The data extraction sheets supported the reading, analysis and synthesis of the documents.

*3.5. Stage 5: Discussion of Most-Relevant Findings*

In stage five, we analyzed the main results, which are summarized in Table 3; Table 4 and Figure 1; Figure 2. In the Discussion Section, results highlight the literature review and future research opportunities on system dynamics in public context. The most relevant findings are described in the following section.

**4. Findings**

We reviewed 144 documents on system dynamic and dynamic performance management from 1976 to 2019, highlighting the following information: (a) Bibliometric information—number of publications and citations, journals, authors, keywords and documents; (b) methodological information—analytical or empirical (if empirical, country of analysis and public sector); and (c) document information (topic and model developed for each paper).

a) The first analysis highlights the main bibliometric information from the dynamic performance management literature in the public context (Figure 1). Firstly, it shows the number of documents and citations, gradually increasing from 1976 to 2019, indicating a fair amount of interest from the scientific community for this topic. Analysis of major journals points to a modest difference between prolific journals. However, "System Dynamics Review" and "Journal of the Operational Research Society" have more citations than other journals. "System Dynamics Review" deals with the implementation of system dynamics to societal, technical, managerial, and environmental issues, whereas "Journal of the Operational Research Society" offers practical solutions to operational issues. The more prolific journals exhibited 51 citations for "Technological Forecasting and Social Change", collected without index keywords and authors' keywords. Hence, the key role of these two documents published in "Technological Forecasting and Social Change" by Forrester and Saeed. Forrester published a computer simulation model of social and economic change in the U.S. and Saeed outlined the advantage and limit of "World Dynamics" and the key role of addressing the correct audience and the ability of systems dynamics to provide complex information. Secondly, it highlights the most prolific authors (e.g., Bianchi, Forrester, Skibniewski and Tatari) (Table 3). Forrester was the creator of the system dynamic approach and Bianchi was the creator of the dynamic performance management approach. Although these authors were key players, the number of their citations is not so high and they do not appear in the analysis of the most important documents. In this ranking, the most cited papers are "Analysing the functional dynamics of technological innovation systems: A scheme of analysis", "Looking in the wrong place for healthcare improvements: A system dynamics study of an accident and emergency department" and "Using system dynamics to improve public participation in environmental decisions". Finally, the last bibliometric information regards the most important keywords (Table 3). The first keyword is "system dynamics", and the last are "performance measurement", "performance" and "dynamic performance management". The "performance" keyword is rarely used in this field. This ranking also points to "public policies" (No. 30) and "simulation" plus "computer simulation" (No. 37) as the most used keywords. The keywords such as "common good/s" and "common resource/s" are not used.

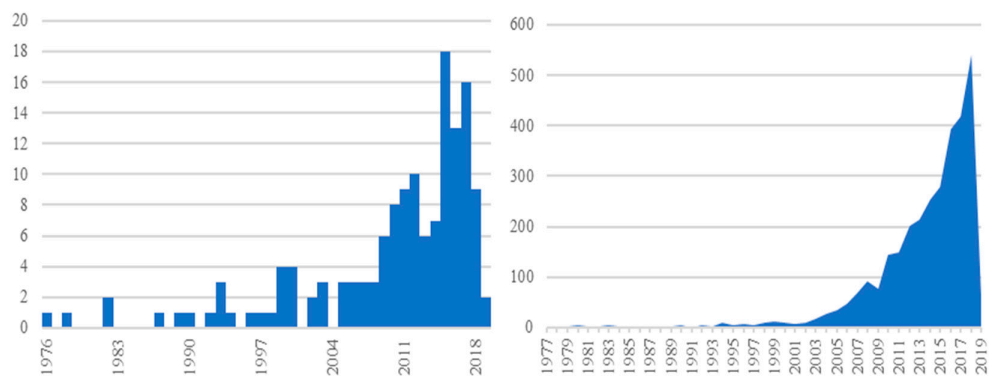

**Figure 1.** Performance bibliometric analysis—Number of documents and citations per year.

**Table 3.** Performance bibliometric analysis—Most Journals, Authors, Documents and Keywords.

| Most Prolific Journal | No. Paper | No. Citations | Main Keywords | No. Paper |
|---|---|---|---|---|
| System Dynamics Review | 16 | 488 | System Dynamics | 89 |
| Systems Research and Behavioural Science | 14 | 133 | System Theory | 51 |
| Journal of The Operational Research Society | 13 | 512 | Public Policy | 30 |
| | | | Computer Simulation | 21 |
| **Most Prolific Authors** | No. Paper | No. Citations | Simulation | 16 |
| Bianchi, C. | 3 | 47 | Decision Making | 13 |
| Forrester, J.W. | 3 | 51 | System Dynamics Model | 13 |
| Skibniewski, M.J. | 3 | 42 | Costs | 12 |
| **Most Cited Papers** | | | | **No. Citations** |
| Analyzing the functional dynamics of technological innovation systems: A scheme of analysis | | | | 639 |
| Looking in the wrong place for healthcare improvements: A system dynamics study of an accident and emergency department | | | | 154 |
| Using system dynamics to improve public participation in environmental decisions | | | | 139 |
| Model calibration as a testing strategy for system dynamics models | | | | 125 |
| How small system dynamics models can help the public policy process | | | | 108 |

*Highlights of first analysis*:

- The scientific community has a fair interest in system dynamics;

- The most prolific journals are "System Dynamics Review" and "Journal of the Operational Research Society", which are specialized in problem-solving;
- The most prolific authors are Bianchi and Forrester, who are the creators of this approach;
- The main keywords are system dynamics, simulation, and public policies. The keywords such as "common good/s" and "common resource/s" are not used.

b) The second analysis shows that 30 documents were based on an analytical methodology and 114 on an empirical methodology (see Figure 2). China, the U.S. and the U.K. are the countries mostly investigating by an empirical methodology (this analysis is referred to the country of the public entity examined), whereas the main public sector using system dynamics is Healthcare, followed by Transport, Government, Waste and Energy respectively.

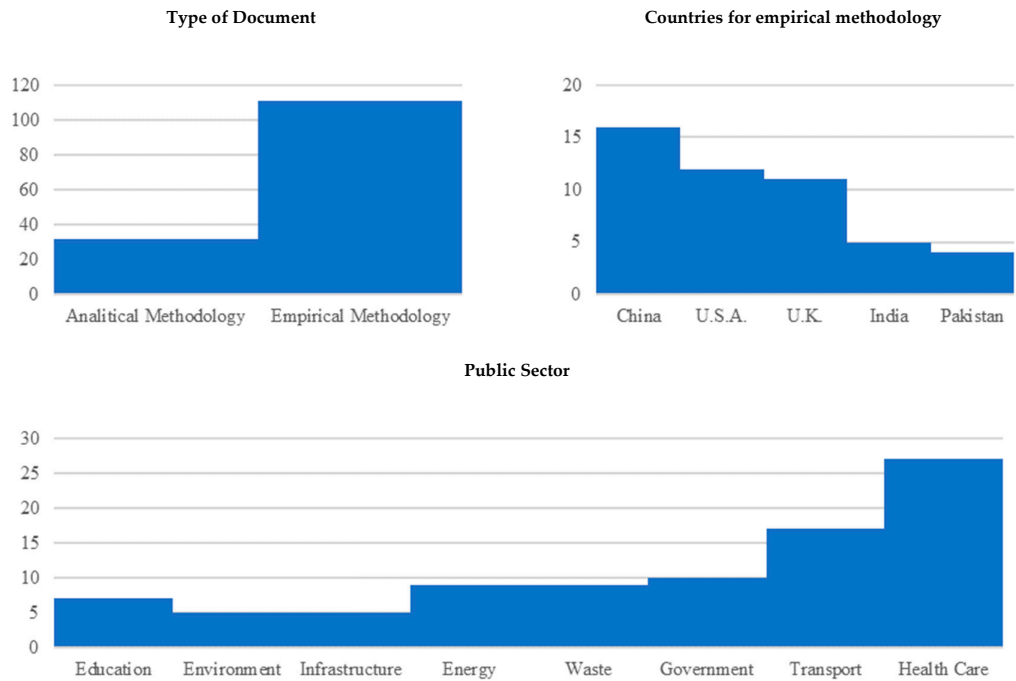

**Figure 2.** Methodology, Country and Sector.

*Highlights of second analysis:*

- Most documents use empirical methodology;
- Major countries investigating include China, the U.S. and the U.K.;
- The main public sector investigating is Healthcare, followed by Transport, Government, Waste and Energy.

c) The third analysis addresses the topic, the model and the common goods analyzed for each empirical paper. Firstly, the analysis of the topics highlights many documents on specific issues, especially Healthcare, for example, HIV, chlamydia and hospital infection, with fewer documents on national and global issues and on more factors analyzed together, for example, social, economic and environmental factors. The documents are based on the problem-solving approach and the holistic view was rarely used, except for some studies on green traffic, pollution and $CO_2$. Secondly, the models developed for the documents highlight various applications of system dynamics, although rarely aimed at performance measurement and management. These models are a dynamic simulation of actual problems and allow forecasting of future developments, for instance, in urban expansion [57] and waste management [58]. These documents rarely adopted the common goods theory and organizational learning. They are often based on system thinking/system dynamics theory. Thirdly, the documents rarely define common goods as a resource that is available for collective use by a community, and whose value and availability can only be maintained and/or developed through the collaboration of the beneficiaries [1]. They adopt implicitly or explicitly the traditional definition of a common good, that is, a resource that is available for collective use, or partially the latest management definition, that is, a resource that is available for collective use by a community, and whose value and availability can only be maintained and/or developed by the community's users. The focus of these documents is toward public goods rather than common goods. However, they do not consider the need for the collaboration of the beneficiaries for (re)generate common goods. This analysis also highlights the lack of studies on common goods created in recent years. For instance, Wikipedia, Wikihow, Open Street Map, Wikiloc, etc.

**Table 4.** Topic, model and issues dealt for each empirical paper.

| Topics and Goods | Models and Issues Dealt |
|---|---|
| National model [8] | interrelated issues of inflation, unemployment, recession, balance of payments, energy, and environment |
| Public policy and rural poverty [53] | policies for alleviating poverty |
| Public policy [54] | policies that address the problems of poverty and hunger |
| Public policy communication [59] | public policies without jargon, mathematics and loop diagrams, but with the art of public communication |
| HIV and AIDS [60] | spread of AIDS that captures complex virological and behavioural features of the epidemic |
| Public school education [61] | education in public schools |
| Community care [62] | community care that combines culture, power and politics to affect the behaviour of a process |
| Pre-college education [63] | public school at all levels for giving cohesion and motivation to public school education |
| Citizen movements [64] | citizen movements and social innovation, providing corrective feedback and creative approaches to evolution |
| Transportation planning [65] | testing of alternative transport-related policies |
| Energy policy [66] | two different energy policy contexts |
| Sustainable urban solid waste [67] | urban solid waste management, which captures dynamic nature of interactions among various components |
| Oil revenues [68] | government financial structure and public services in course of development based on oil revenues |
| AIDS [69] | AIDS incubation time distribution |
| Improving SD [70] | dynamic optimization |
| Public utility market [71] | public utilities in the energy market |
| Waiting lists in public hospital [72] | waiting lists in public hospital |
| Waiting lists in public hospital [73] | delays of accident and emergency in public hospitals |
| Chlamydia [74] | screening program of Chlamydia |
| Development project [75] | development projects |
| Organizational behaviours [76] | organizational behavioural setting in the infrastructure sector |
| Public decision-making [77] | support of stakeholder advisory group examining transportation and related air quality problems |
| Benchmark design [78] | benchmarking in education to identify the gaps in performance |
| Defense technology [79] | development of defense technology |
| Testing strategy [80] | testing strategy |
| Chlamydia infection [81] | capturing chlamydia infection within a population |
| Off-Shoring IT work [82] | offshoring IT-Word and evolution beyond current high-growth period |
| Feedback in health care [83] | clinical guidelines to demonstrate the potential benefits of changing the goals that drive activity |
| Road congestion [84] | road congestion with alternative policies, which promote public transport |
| Public policy in tobacco [85] | price influence on the use and consequences of tobacco |
| Chlamydia screening [86] | infection and the cost-effectiveness of the intervention strategies of chlamydia screening program |
| Climate change [87] | global policy of climate change |
| Sustainable development [88] | monitor for sustainable development of the urban water system |
| Solid waste management [89] | different pricing systems within the real world of solid waste management |
| Public expenditure [90] | alternative policy to pursuit pro-growth |
| Resource management [91] | context-dependency of participatory processes in natural resource knowledge management |
| Public expenditure [92] | impact of public expenditure on human development and economic growth |
| Public sport service [93] | understanding of customer loyalty in a public sports service |
| Vaccination policy [94] | different decision rules with respect to vaccination policy for eradicable disease |
| Waste management system [95] | development of a solid waste management system |
| Public health websites [96] | understanding of web-support for public communication in complex area of healthcare |
| Strategic conflict [97] | strategic of modern conflict |
| Public policy [98] | learn from public health problem |
| Organizational change [10] | understanding of relationships between political and organizational system in public sector |
| Accountability public sector [99] | impact of back and front office units on a public sector organization's performance drivers |
| Participatory decision-making [100] | participatory approach in environmental decision-making |
| Epidemics [101] | dynamic transmission of tuberculosis, drug-resistant tuberculosis, HIV and human immunodeficiency virus |
| Income distribution [102] | model to detect persistent asymmetric income distribution |
| Globalization [103] | unearthing the dynamic processes underlying globalization |
| Economic growth [104] | relationship between economic growth, complexity, "maturity" of the population |
| Public policy [105] | lessons for policymaking stemming from the endogenous and aggregate perspective |

| Topics and Goods | Models and Issues Dealt |
|---|---|
| Natural-societal risk [106] | stratospheric ozone to identify points to influence policymakers in addressing risks in a natural system |
| Waste management [107] | cost-benefit of construction and demolition waste management practices |
| Strategic conversations [108] | soft operational research strategy making an intervention with a top management team |
| Interactive knowledge management [109] | knowledge management for improving quality of knowledge in non-hierarchical policy-making groups |
| Waste management [110] | economic feasibility of recycling and ratio of savings to costs in construction and demolition waste management |
| Emergency service [111] | Supply–demand equation that delivers frontline policing services |
| Air transportation [112] | policies and strategies for air transportation (e.g., policies for reducing emissions of commercial aviation |
| Local public policy [113] | local public policy process |
| Public sector unit [114] | mortality rate of public sector units |
| Concession pricing model [115] | determining a rational concession price for highway projects based on pro forma financial statements |
| Highway sustainability [116] | highway sustainability based on individual policies with a crucial impact on the success of policymaking |
| Academic performance [117] | factors that affect academic performance of migrant student residing in poor migrant neighborhoods |
| Maritime Sustainability [118] | integrated sustainability assessment process |
| Information in health care [119] | emergent market behaviour enabling evaluation of different public health policies |
| Climate change [120] | electricity industry at the same time as addressing climate change issues |
| Participatory urban [121] | urban sensing for improving municipality's availability of environmental information at comparable cost level |
| Wireless local area [122] | effect of subsidization, revenue sharing, and alliance strategies |
| Airport competitiveness [123] | airport investment and city R&D inputs, which support decision-makers on airport competitiveness |
| Electronic health records [124] | combination of measures can promote the adoption of electronic health records by different stakeholders |
| Housing market [125] | house prices, balances between supply and demand, construction companies' earnings, and vacancy rates |
| Participatory policy [126] | participatory procedures to the governance of wider public policy issues |
| Debt accumulation [127] | public debt in developing countries and its two-way linkages with economic growth and public finances |
| Evaluation of public projects [128] | future social benefits and costs of public projects (financed especially from funds of the European Union) |
| Pension risk management [129] | dynamics simulation in analyzing pension expenditure in evaluating impact of policy decisions |
| Green building [130] | model for promoting the green building market |
| Innovation systems [131] | innovation race: competitiveness and influence through innovation; the stakes of global public goods |
| Waste management [132] | impacts of two alternatives for the management of construction and demolition waste |
| Community-based prevention [133] | implications for community-based prevention marketing's developers and other social marketers |
| Bodyweight gain/loss [134] | health care professionals failed to understand simple dynamic impact of energy intake and energy expenditure |
| Carbon labelling [135] | consumers' responses to carbon labelled products and number of consumers |
| Societal ageing [136] | demographic and social security system to investigate governmental retirement and healthcare contributions |
| Incineration projects [137] | determining concession period and subsidy for build-operate-transfer waste-to-energy incineration projects |
| Cyberattacks [138] | several cyber-attack scenarios |
| Energy and $CO_2$ emissions [139] | energy consumption and $CO_2$ emission intensity in China from 2013 to 2020 via SD simulation |
| National healthcare [140] | national health-care system, validated by a simple computational prototype |
| Ridesharing services [141] | dynamic ridesharing platform that should operate mostly in urban areas |
| Truck weight regulations [142] | long-term effects of alternative truck weight regulation policies on sustainability of highway transportation |
| Environmental accumulation [143] | building an operational understanding of accumulations and suggest design considerations |
| Performance-based payment [144] | impacts of performance-based payment system on the behaviours of public hospitals and their physicians |
| HIV testing law [145] | HIV testing and care system to help administrators |
| Climate negotiations [146] | evaluations of the impact of world climate with diverse group |
| Hip replacement and obesity [147] | influence of obesity on the healthcare expenditure in the domain of hip implants |
| Post-disaster recovery [148] | understanding overall recovery efforts in the whole region from a holistic perspective |
| Stakeholders satisfaction [149] | interests of the principal stakeholders in Public-Private Partnerships- PPP |
| Risk assessment of PPP [150] | demand risk in road projects |
| Investment projects [151] | feasibility analysis for public investment projects |
| Cost-sharing in networks [152] | directed networks with positive externalities induced by cost-sharing |
| Client perceptions [153] | outcomes of potential clients in public sector |
| Pension system [154] | policy reforms of changing contribution rates and operating performance of pension system |
| Addressing conflicted situation [155] | ambiguous and conflicted situation |
| Public R&D institute [156] | process analysis of industrial technology research institute |
| Green traffic system [157] | ownership of the private car, the increase of vehicle exhaust stock and environmental pollution |
| Transportation costs [158] | transport-related costs and develop ways to decrease these costs |
| Green economy [159] | green transition system, suggesting learning and adaptive mechanisms involving stakeholders |
| Pain medicine prescribing [160] | impact of three types of policy interventions to reduce adverse outcomes in pain medicine prescribing |
| Child protection system [161] | systems thinking ideas in child protection |
| Innovative entrepreneurship [162] | innovative entrepreneurship in business cycle scenarios |
| Carbon footprint reduction [163] | public transportation to reduce $CO_2$ emissions and energy consumption and to increase roadway safety |
| Infrastructure development [164] | infrastructure development by integrating technical, economic and operational aspects |
| Natural gas generation [165] | natural gas power generation |
| Parking policy [166] | parking policies |
| Childhood immunization [167] | public health to examine the childhood immunization system to identify pathways of delivering vaccination |
| Children overweight [168] | micro (family) and macro (governmental policies) levels that need for reducing obesity and overweight |
| Eco-friendly vans [169] | eco-friendly vans for urban freight distribution |

| Topics and Goods | Models and Issues Dealt |
|---|---|
| Decision support [170] | decision support system which indicate what issues and stakeholders are to be included in decision analysis |
| Urban development structure [171] | mechanisms contributing to the low public transportation use |
| E-waste generation and recycle [172] | prediction of e-waste generation and distribution through different disposal pathways |
| Dynamic performance management [55] | an outcome-based approach to solving wicked policy problems |
| Work-education mismatch [173] | the education-workforce pipeline |
| Socio-technical transition in elec. [174] | impacts and side effects of stable feed-in tariffs of wind and solar electricity |
| Carbon reduction labelling [175] | government of possible sustainable policy design to promote low-carbon development |
| Emerging research institutions [176] | optimizing budget resource planning for sustainable supply chain networks and other institutions |
| Alternative fuel vehicles [177] | impact of public policies in the long-term diffusion dynamics of alternative fuel vehicles in Brazil |
| Population health [178] | national systems thinking approach towards improving the health of communities and populations |
| Renewable energy power [179] | long-term development of the renewable energy power industry |
| Highway projects [180] | interactions among the highway pavement performance and stakeholders' relevant factors |
| Risk management [181] | impact of maritime regulatory changes |
| Waste reduction management [182] | construction professionals for waste reduction measures |
| Community involvement [183] | collaborative networks including local communities |
| Strategic workforce planning [184] | healthcare workforce planning to support strategic planning with multiple methodologies and modelling |
| Traffic and emissions impact [185] | advantages of two policies, which reduces the emission of vehicle pollutants and alleviate traffic congestion |
| Essential medicines stock [186] | medicine supply chain in rural parts of a developing country |
| Electric vehicle charging [187] | charge operators' profits and reduce charge price |
| Technology park [188] | investment of the development of a technology park |
| Childhood overweight [189] | government and stakeholders responsible for meeting the target of reducing childhood overweight |
| Renewable energy [190] | evolution from feed-in tariff to renewable portfolio standards scheme, influenced by renewable and strategies |

*Highlights of third analysis:*

- The documents aim to solve specific problems, especially Healthcare, rather than national and global problems;
- The documents are based on the problem-solving approach without a holistic view;
- The models rarely deal with performance measurement and management.
- The documents rarely show the need for the collaboration of the (re)generation of common goods.
- This document highlights the lack of studies on common goods created in recent years (e.g., Wikipedia).

**5. Discussion**

Since the first publication, system dynamics has been recognised as a useful approach for including the complexity of a public context to identify the economic, environmental and social change in behaviour and policy of a national model [8]. Recently, the literature has highlighted an innovative system dynamics approach for dynamic performance management [9]. Although recognised as an innovative approach for managing common goods and for linking the management of common goods and organizational performance, this approach is still rarely adopted for this purpose in the public context.

We systematically investigated the literature on system dynamics and dynamic performance management in the public context to show (a) the state of the art and (b) future opportunities on the use of system dynamics in public context.

*a)* The findings point to an interest by the scientific community in system dynamics of public context in terms of the number of documents and citations. The more popular journals are specialized in problem-solving and the most prolific authors, the very creators of this approach, received few citations. The results also indicate scant adoption of this approach in performance management, as highlighted by keywords and models (see Table 3 and Table 4). The focus of these studies, which are mainly empirical in countries such as China, the U.S. and the U.K., is often the computer simulation of specific problems in healthcare. Although many of them deal with specific issues, some studies describe global models on issues such as climate, waste, and energy. In many cases, these models show the management of public goods; however, these models rarely deal with common goods created recently (e.g., Wikipedia, Wikihow) and address the linking between common goods and organizational performance. They do not integrate all elements through the holistic view, lacking specific focus on organizational performance. The state-of-the-art as regards dynamic performance management of the public context can be considered embryonic. Although system dynamics pertains to specific and operational issues, its application lacks a global and holistic vision. In 1982, Forrester stated that a model had to be global or national (not merely regional), draw heavily on mental and not just written and numerical databases, and have time horizons of perhaps 100 years. It had to include social and economic change and provide insights in behaviour and policy at each stage of development. Application and interpretation of a model had to occur in cumulative fashion from testing of individual sectors to combined testing of two or more sectors and, finally, policy studies in the complete assembled national model [8]. However, although the theory on system dynamics was clearly oriented toward a global and holistic vision, in practice the trend is toward managing specific public goods, the vision shifting from holistic to fragmented, and moving away from the possibility of adopting this approach for inclusive and dynamic performance management. Furthermore, it rarely considered common goods as a resource that are available for collective use by a community, and whose value and availability can only be maintained and/or developed through the collaboration of the beneficiaries [1]. The focus of these documents is toward public goods rather than common goods. They do not highlight the need for the collaboration of the beneficiaries to (re)generate common goods. Thus, it lacks studies on the recent common goods like Wikipedia and Wikihow. *To conclude, the state of art still is far from being of use in supporting dynamic performance management in the public context, its "embryonic status" remaining a great challenge. It is stopped on the economical and legal vision of public and common goods, that is, goods that do not need the collaboration of the beneficiaries.*

b)　This literature review outlines future opportunities for dynamic performance management in the public context. The main opportunities imply a need for further investigations in the management of common goods and organizational management. In specific, the main opportunities regard the adoption of system dynamics in performance management, decision-making, and sustainability. Overall opportunities regard the need for studies to improve the theoretical background of this approach and the knowledge on the common goods created recently. The first opportunity covers studies on design, implementation and adoption of system dynamics in performance management. Results highlight a fair interest by the scientific community in system dynamics of public context, but also inadequate adoption of this approach in performance management, as indicated by the keywords and models. New models should be developed to improve performance measurement and management. The second opportunity regards the possibility of new research on global models to predict the social, environmental and economic effects of new policies. The assessment of economic, environmental social and policies should be tackled from a holistic perspective, using methodologies such as dynamic system models [191]. Although in theory a system dynamic model should be at least national in scope, the literature is often focused on sectoral problems, for example, infection in a hospital or pollution in a city. These models often lack the holistic aspect. New models should be developed looking at national and international policies and their impact in terms of social and economic effects. Furthermore, they should provide insights into behaviour and policy at each stage of development. The third opportunity concerns new studies on the integration of new and integrated topics in system dynamics. Often the focus of these studies is computer simulation; however, this approach plays a key role in different ways. New models should study the decision-making process, explain better the sustainable and the social and economic effects of public policies. Overall academic opportunities point to the need for studies to improve the theoretical background of this approach. In fact, the literature does not comprise sufficient theoretical studies to improve the approach in terms of strategy implementation, forecasting of effects and performance measurement, that is, past, present and future. Another overall opportunity is the study of common goods as a resource that is available for collective use by a community, and whose value and availability can only be maintained and/or developed through the collaboration of the beneficiaries [1]. The public context needs research on common goods rather than public goods. The literature urgently needs research on the common goods created recently. *To conclude, overall opportunities call for a strong theoretical background on dynamic performance management and more research effort on:*

- *Design, implementation, and adoption of system dynamics in performance management;*
- *Global forecasting models of the social and economic effects of new policies;*
- *Decision-making and sustainability;*
- *The use of the dynamic performance management approach in the public context;*
- *And especially, recent common goods (e.g., Wikipedia), according to the last management definition.*

Finally, the dynamic performance management approach should be more user-friendly. Its understanding is still a great challenge. In fact, model design is not widely accessible because of the complexity of the approach and the highly professional skills needed to learn its application. Currently, it is accessible only to experts in system dynamics and software simulation. Complexity is a big limit to its adoption.

## 6. Conclusions

Since the publication of "The system dynamics national model: Understanding socio-economic behaviour and policy alternatives" on "Technological Forecasting and Social Change", system dynamics has been recognised to be a useful approach for including the complexity of a public context to identify economic, environmental and social change about behaviour and policy of a national model. Although system dynamics is recognised to be an innovative approach to improving performance management, it is as yet rarely adopted in the public context.

We systematically reviewed the scientific literature of system dynamics and dynamic performance management in the public context.

The literature of system dynamics and dynamic performance management is as yet embryonic, highlighting a fair number of documents and citations involving system dynamics in the public context and inadequate adoption of this approach in managing common goods and organizational performance. Findings point to mainly empirical studies in countries such as China, the U.S. and the U.K., focused on computer simulation of specific healthcare issues. Although various studies deal with specific problems, some describe global models for some issues, including climate, waste, and energy. These models often anticipate social, environmental and technological changes, but rarely explain changes entirely, in that they often lack a holistic view and do not include a specific focus on common goods, missing the focus on the collaboration of the beneficiaries to (re)generate common goods and on the performance measurement and management.

The theory on system dynamics is clearly oriented toward a global and holistic vision, but in practice the trend is toward solving specific issues, with a shift from a global to a fragmented vision, leading away from the possibility of adopting this approach for dynamic performance management.

Even if potentially useful for dynamic performance management, system dynamics as yet lacks powerful theoretical and practical foundations for its adoption in performance management in the public context.

This research project is meant as a theoretical and practical contribution. Firstly, it offers scholars research opportunities for the study of this approach to improve system dynamics in the public context. Overall academic opportunities imply (a) a need for a strong theoretical background on dynamic performance management approach and more research on design, implementation and adoption of system dynamics in managing common goods and organizational performance, and decision-making. (b) Then, they imply an effort on the study of common goods defined as a resource that is available for collective use by a community, and whose value and availability can only be maintained and/or developed through the collaboration of the beneficiaries [1]. The public context needs research on common goods rather than public goods. They urgently need research on recent common goods like Wikipedia and Wikihow.

Secondly, this review mentions the main documents from which to obtain useful ideas on how to replicate system dynamics in other similar public contexts. For example, Bianchi et al. illustrate the adoption of the system dynamic approach in three different public organizations (public utility, healthcare organizations and municipal theatre), describing how system dynamics facilitates an understanding of the relationships between the political and organizational systems in the public sector and how it improves the efficiency and the end results, in spite of their constraints.

The limitations of this study regard mainly the research criteria and the analysis of topic and model for each paper.

Firstly, notwithstanding the inclusion of three subject areas, that is, "Business, Management and Accounting", "Economics, Econometrics and Finance" and "Decision Sciences", some other areas could be missing, e.g., "Social Science", "Engineering" and "Computer Science". However, the areas chosen should adequately focus the research, that is, managerial problems versus performance problems. Secondly, the analysis of topic and model for each paper look like a sort of "tag" of the main topic of the paper and lack a precise spotting of which is the "common good" considered. However, thanks to this analysis, we provide an overall insight for the management of common goods and organizational performance and show scholars the main research opportunities and useful ideas on the adoption of system dynamics for managing common goods.

Although these limitations may represent potential weaknesses, they also allowed us to focus on common goods and performance management in a wider sense. They also became the strong point of the research. Their use in selecting papers permitted the inclusion of focused contributions of specific research streams and could, therefore, support the identification of a database for fully developing the system dynamic method. *In conclusion, the great challenge for scholars will be to translate system dynamics from an approach for experts to a user-friendly system for everyone, including the latest management definition of common goods. Meeting this challenge will lead to a worldwide application of system dynamics, also in managing common goods and organizational performance.*

**Author Contributions:** Introduction, A.S. and E.S.; Literature Background, A.S. and E.S.; Methodology, A.S. and E.S.; Findings A.S. and E.S.; Discussion A.S. and E.S.; Conclusions A.S. and E.S.

**Funding:** This research received no external funding.

**Conflicts of Interest:** The authors declare no conflicts of interest.

*Sustainability* **2019**, *11*, 6435 18 of 21

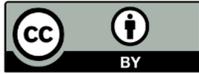